%% file: main.tex
\documentclass[10pt,journal,compsoc]{IEEEtran}

\usepackage[switch]{lineno}
\AtBeginDocument{%
  \providecommand\BibTeX{{%
    \normalfont B\kern-0.5em{\scshape i\kern-0.25em b}\kern-0.8em\TeX}}}

\usepackage[svgnames]{xcolor}
\usepackage{url}
 \usepackage{graphicx}
 \graphicspath{ {images/} } %
\usepackage[inline]{enumitem}
\usepackage{xurl}
 \usepackage{tikz}
 \usepackage{tabularx}
 \usepackage{booktabs}
\usepackage{makecell}
\usepackage{array}
 \usepackage{algorithm}
 \usepackage{algorithmic}
 \usepackage{comment}
 \usepackage{caption}
\usepackage{subcaption}
\usepackage{tcolorbox}
\usepackage{circledsteps}
\usepackage{multirow}
\usepackage{pgfplots}
\usepackage{pgfplotstable}
\usepackage{ragged2e}
\usepackage{cite}
\usepackage{tablefootnote}

\pgfplotsset{compat=1.18}
\newcolumntype{S}{>{\hsize=.35\hsize \arraybackslash}X} 
\newcolumntype{B}{>{\hsize=.65\hsize \arraybackslash}X} 

\usetikzlibrary{arrows,positioning,shapes.geometric}
\usepackage{tcolorbox}
\usepackage[multiple]{footmisc}
\begin{document}
\title{\Large \bf \textit{Trusting code in the wild}: Exploring contributor reputation measures to review dependencies in the Rust ecosystem}

\author{Sivana~Hamer, Nasif~Imtiaz, Mahzabin~Tamanna, Preya~Shabrina,
        and~Laurie~Williams%
\IEEEcompsocitemizethanks{\IEEEcompsocthanksitem Sivana Hamer, Nasif Imtiaz, Mahzabin Tamanna, Preya Shabrina, and Laurie Williams are with the Department of Computer Science, North Carolina State University, Raleigh, NC, 27606.\protect\\
E-mail: \{sahamer, simtiaz, mtamann, pshabri, lawilli3\}@ncsu.edu}%
}

\input{sections/_commands}
\input{sections/0.abstract}

\maketitle

\input{sections/1.introduction}
\input{sections/2.background-rw}

\input{sections/3.0.methods}
\input{sections/3.1.network}

\input{sections/3.2.survey}

\input{sections/3.3.model}

\input{sections/4.RQ1}

\input{sections/5.RQ2}

\input{sections/6.discussion}

\input{sections/7.limitation}

\input{sections/8.conclusion}
\input{sections/9.acknowledgments}

\bibliographystyle{IEEEtran}
\bibliography{bibliography}
\end{document}

%% file: sections/_commands.tex
\newcommand\sivanatodolabel[3]{\textcolor{#1}{\textbf{Sivana pending #2: }{#3}}}
\newcommand\sivanarestructure[1]{\sivanatodolabel{LightSkyBlue}{restructure}{#1}}
\newcommand\sivananovelty[1]{\sivanatodolabel{MediumPurple}{novelty}{#1}}
\newcommand\sivanaimplications[1]{\sivanatodolabel{LightCoral}{implications}{#1}}
\newcommand\sivanaanalysis[1]{\sivanatodolabel{Crimson}{analysis}{#1}}
\newcommand\sivanaclarify[1]{\sivanatodolabel{Goldenrod}{clarify}{#1}}
\newcommand\sivanapending[1]{\sivanatodolabel{MediumVioletRed}{all}{#1}}
\newcommand\sivanalong[1]{\sivanatodolabel{DarkOrchid}{overmotivated/long}{#1}}
\newcommand\sivanafinal[1]{\sivanatodolabel{MediumVioletRed}{final qualitative coding}{#1}}
\newcommand\sivanalastnumber[1]{\textcolor{red}{\textbf{#1}}}

\newcommand\boxrecommendation[2]{\begin{tcolorbox}\textbf{Recommendation #1:} #2\end{tcolorbox}}

\definecolor{blue1}{RGB}{152,209,209}
\definecolor{blue2}{RGB}{13,141,142}

\definecolor{pink1}{RGB}{223,151,158}
\definecolor{pink2}{RGB}{200,0,100}

\definecolor{gray1}{RGB}{218,218,218}

\newcommand\distributionsurvey[8]{
\resizebox{0.08\linewidth}{4pt}{{#7}}
\resizebox {0.81\linewidth} {6.5pt} {%
\begin{tikzpicture}
    \begin{axis}[
          axis background/.style={fill=gray!30, draw=gray!30},
          axis line style={draw=none},
          tick style={draw=none},
          ytick=\empty,
          xtick=\empty,
          ymin=0, ymax=0.70,
          xmin=0, xmax=6
        ]
        \addplot [ybar interval=.5,fill=blue2,draw=none,] coordinates {(6*#1,1) (0,0.30)}; %
        \addplot [ybar interval=.5,fill=blue1,draw=none,] coordinates {(6*(#1+#2),1) (6*#1,1)}; %
        \addplot [ybar interval=.5,fill=gray1,draw=none,] coordinates {(6*(#3+#2+#1),1) (6*(#2+#1),1)}; %
        \addplot [ybar interval=.5,fill=pink1,draw=none,] coordinates {(6*(#4+#3+#2+#1),1) (6*(#3+#2+#1),1)}; %
        \addplot [ybar interval=.5,fill=pink2,draw=none,] coordinates {(6*(#5+#4+#3+#2+#1),1) (6*(#4+#3+#2+#1),1)}; %
    \end{axis}%
\end{tikzpicture}%
}
\resizebox{0.08\linewidth}{4pt}{{#8}}

}

\newcommand\legendtable[2]{
\resizebox {0.02\linewidth} {6.5pt} {%
\begin{tikzpicture}[]
\begin{axis}[
      axis background/.style={fill=white!30, draw=white!30},
      axis line style={draw=none},
      tick style={draw=none},
      ytick=\empty,
      xtick=\empty,
      ymin=0, ymax=0.70,
      xmin=0, xmax=6]
\addplot [ybar interval=.5,fill=#2,draw=none,] coordinates {(4.5,1) (0,0.30)}; %
\end{axis}%
\end{tikzpicture}%
}%
#1
}

%% file: sections/0.abstract.tex
\IEEEtitleabstractindextext{
\begin{abstract}

Developers rely on open-source packages and must review dependencies to safeguard against vulnerable or malicious upstream code.
A careful review of all dependencies changes often does not occur in practice. Therefore, developers need signals to inform of dependency changes that require additional examination.
The goal of this study is to help developers prioritize dependency review efforts by analyzing contributor reputation measures as a signal.
We use network centrality measures to proxy contributor reputation using collaboration activity.
We employ a mixed method methodology from the top 1,644 packages in the Rust ecosystem to build a network of 6,949 developers, survey 285 developers, and model 5 centrality measures.
We find that only 24\% of respondents often review dependencies before adding or updating a package, mentioning difficulties in the review process.
Additionally, 51\% of respondents often consider contributor reputation when reviewing dependencies.
The closeness centrality measure is a significant factor in explaining how developers review dependencies.
Yet, centrality measures alone do not account for how developers choose to review dependencies.
We recommend that ecosystems like GitHub, Rust, and npm implement a contributor reputation badge based on our modeled coefficients to aid developers 
in dependency reviews.
\end{abstract}
}

%% file: sections/1.introduction.tex
\section{Introduction}

Modern software relies on open-source packages in its supply chain.
However, the dependence on open-source packages has opened up new attack vectors as vulnerable and malicious code can infiltrate, propagating to downstream dependents~\cite{ohm2020backstabber}.
An example of such an attack is the xz backdoor.
\textit{xz-utils} is a data compression library used, among others, in Linux distributions~\cite{what2024}.
In March 2024, 
malicious commits in \textit{xz-utils} introduced a backdoor affecting SSH as a dependent~\cite{freund2024}. 
The United States Cybersecurity and Infrastructure Security Agency (CISA) and Red Hat released a security advisory due to the severity of the attack~\cite{cisa2024, redhat2024}.

The software supply chain community has recommended security safeguards against attacks, such as 
\textit{reviewing the package code before adding or updating a dependency}~\cite{imtiaz2022open,yang2021solarwinds}.
Selecting a trustworthy and audited dependency version can safeguard the dependent.
For the upstream, developers should \textit{carefully review incoming changes}~\cite{goyal2018identifying, ladisa2023sok}.
If no vulnerable or malicious code is introduced in dependencies, dependents remain unaffected.
Practitioners have, consequently, incorporated reviews in software supply chain
solutions.
Examples include Supply chain Levels for Software Artifacts (SLSA)~\cite{slsa}, OpenSSF Scorecard~\cite{scorecard}, and cargo-crev~\cite{cargocrev}.

However, reviewing all changes within the software supply chain often does not occur in practice and is difficult for developers.
Only $30\%$ to $34\%$ of packages within open-source ecosystems review their code~\cite{zahan2023Openssf}.
Even when developers review, only $11\%$ of package updates have all code reviewed~\cite{imtiaz2023Your}.
For the same reason, in a survey of 134 developers and 17 security experts, Ladisa et al.~\cite{ladisa2023sok} found that reviews are the best software supply chain safeguard but are among the most costly~\cite{ladisa2023sok}.
Although research has studied how developers choose dependencies~\cite{larios2020selecting, mujahid2023characteristics, tanzil2024people} and review code within their projects~\cite{ford2019beyond,tsay2014influence,zhang2022pull}, less attention has been given to proposing measures that can be used as signals for dependency review.
Therefore, developers need measures to identify dependency changes requiring additional examination.

\textbf{The goal of this study is to help developers prioritize dependency review efforts by analyzing contributor reputation measures as a signal.}
To that end, we study with mixed methods the Rust ecosystem —that is— the packages hosted on Crates.io~\cite{cratesio}. 
We leverage a developer social network~\cite{herbold2021systematic} to calculate centrality measures as a proxy for contributor reputation~\cite{bosu2014impact} within the network.
Hence, we construct a social network of developers from the most downloaded 1,644 Rust packages based on file co-edition and author-reviewer collaboration activity.
We then gathered the associated developer network centrality measures.
We investigate the following research questions:

\begin{quote}
    \textbf{RQ1:} 
    How do developers choose to review dependencies in the Rust Ecosystem? 

\end{quote}

Towards RQ1, we surveyed 285 Rust developers to understand what strategies developers use, how developers recognize other members within the community, and how developers review upstream changes based on contributor identity.
We quantitatively and qualitatively analyze the responses.
Among our findings, only 24\% of respondents reviewed dependencies, mentioning reviewing all dependencies changes as infeasible.
Thus, developers mentioned employing strategies to reduce the review level.
The highest-rated factor considered by 51\% of respondents for dependency reviews was contributor reputation.

\begin{quote}
    \textbf{RQ2:} How can network centrality measures, as proxies for contributor reputation, signal the need to review dependencies?
\end{quote}

Towards RQ2, we employed multivariate mixed-effect linear regression models to determine the effectiveness of network centrality measures in predicting the level of dependency review. 
We find that the centrality measure, closeness centrality, was a statistically significant variable across our models, explaining the level of review for a dependency.
Still, at most, the centrality measures could account for 13\% of the models' variation.
Hence, though centrality measures can serve as a signal for dependency review, developers also use other non-contributor reputation factors.

We provide recommendations for ecosystems, developers, and researchers based on our findings.
Among our recommendations, ecosystems should provide a badge indicating reputed contributors in the community.
Ecosystems can calculate contributor reputation using network centrality measures and our modeled coefficients.
With a standardized ecosystem-level badge, we can support developers in dependency reviews in the software supply chain.

\textbf{In summary, our contributions are:} 
\Circled{1} a network centrality measure solution to help prioritize dependency review efforts; 
\Circled{2} empirical insights on how developers choose to review dependencies with an available codebook;
\Circled{3} models to predict dependencies' level of review using network centrality scores;
\Circled{4} a list of role-based recommendations from our findings to guide developer dependency review efforts; and
\Circled{5} empirical insights on the developer community structure through a social network of Rust with an available collaboration activity dataset.

The Institutional Review Board (\textit{IRB}) approved our survey protocol. 
Our codebook and dataset are available in the supplemental material~\cite{us2024Supplemental}. %
The rest of the paper is structured as follows: 
Section~\ref{relwork} discusses the background and related work. 
Section~\ref{sec:method} explains our mixed-methods methodology.
Afterward, Section~\ref{sec:rq1} and \ref{sec:rq2} present the findings for our two research questions. 
Section~\ref{sec:discussion} discusses the implications and recommendations. 
Section \ref{limitation} details the limitation of our measure and threats to the validity of our study before we conclude in Section~\ref{conclusion}.

%% file: sections/2.background-rw.tex
\section{Background and Related Work}
\label{relwork}

In this section, we explain the key concepts of our study and discuss the related work. 

\subsection{Software Supply Chain}\label{bac:ssc}
Modern software extensively uses open-source packages. 
Each major programming language has a registry supplying freely available packages, such as Crates.io for Rust, npm for JavaScript, PyPI for Python, and RubyGems for Ruby. 
Developers in these languages can use packages and benefit from code reuse~\cite{feitosa2020code}. 
When projects include open-source packages in their codebase, the package becomes an \textbf{\textit{upstream dependency}}. 
Meanwhile, the packages that include the upstream dependency are \textbf{\textit{downstream dependents}}.
A developer serves both as a downstream user when using dependencies and an upstream developer when contributing to a package others may use.

While package registers benefit code reuse, packages also come with security risks. 
Open-source packages are at risk of software supply chain attacks~\cite{ohm2020backstabber, ladisa2023sok}. 
Particularly in the context of our study, vulnerable or malicious code from dependencies.
The risk of known and unknown vulnerabilities in dependencies has been studied in the literature~\cite{imtiaz2022open, decan2018impact, lauinger2018thou, imtiaz2021comparative}. 
Research has also investigated malicious packages~\cite{zahan2022weak, sejfia2022practical, ladisa2022towards} and commits~\cite{gonzalez2021Anomalicious}.

Reviewing the package code before adding or updating a dependency~\cite{imtiaz2022open,yang2021solarwinds} and carefully reviewing incoming changes~\cite{goyal2018identifying, ladisa2023sok} are recommended safeguards. 
Consequently, industry framework solutions to review dependencies have emerged. 
Supply chain Levels for Software Artifacts (SLSA)~\cite{slsa}, a security framework for using open-source packages, requires two trusted actors to review all code changes to achieve the highest security rating.
OpenSSF Scorecard~\cite{scorecard}, a security health metrics tool for open-source, integrates within their source code risk assessment score if changes are reviewed.
Dependabot~\cite{dependabot} monitors and alerts vulnerabilities in dependencies.
Third-party package auditing toolings, such as cargo-crev~\cite{cargocrev} and cargo-vet~\cite{cargovet}, have also been created.
We contribute to software supply chain review solutions by proposing and analyzing contributor reputation as a signal to complement and strengthen current work.

\subsection{Trust in the Software Supply Chain}

One of the top five software supply chain practitioner concerns is how to trust code developed by others~\cite{enck2022top}.
A lack of trust is cost-ineffective, requiring developers to divert development resources into reducing security risks~\cite{boughton2024Decomposing}.
Therefore, work on understanding trust within the software supply chain has been of research interest.
Research has investigated trust processes and challenges for practitioners using open-source projects~\cite{wermke2022committed, ghofrani2022trust} and how trust is exhibited in GitHub pull requests~\cite{sajadi2023interpersonal}.

Part of trusting code from others also involves determining if someone is trustworthy.
As such, work has started to propose contributor trust measures~\cite{gonzalez2021Anomalicious, boughton2024Decomposing}.
The most similar work is by Boughton et al.~\cite{boughton2024Decomposing}.
They identify trust contracts that interplay between trustors and trustees.
They then decompose trust contracts in (i) the propensity of the trustor to trust, (ii) the perceived integrity of the trustee, (iii) the perceived benevolence of the trustee, and (iv) the perceived ability of the trustee to perform an attack.
The operationalization of the construct is left to future work.
We also go towards measures for contributor trust in the software supply chain.
We analyze network centrality measures as a proxy of contributor reputation, an operalization for the perceived benevolence trust component.

\subsection{Signaling Theory in Open-Source Software}

In coding platforms, signals provide easily observable information to developers, such as repository badges in GitHub~\cite{trockman2018adding}.
Signals within open-source contexts are used for various decisions, including selecting which project to contribute to~\cite{qiu2019signals} and choosing tasks to work on~\cite{santos2022choose}.
When developers review pull requests in their projects, they also consider technical and social indicators of the contributor beyond the code~\cite{ford2019beyond}. 
Research has found that social factors such as contributor reputation, prior interaction, social strength, and community standing affect review outcomes~\cite{bosu2014impact,tsay2014influence,zhang2022pull}.

Signals are used not only upstream but also downstream.
For example, when choosing packages and libraries as a dependent~\cite{wermke2022committed, larios2020selecting, mujahid2023characteristics, tanzil2024people, miller2023We}.
Notably, developers consider contributor reputation a signal when choosing dependencies~\cite{zahan2022weak, miller2023We, holtervennhoff2023Wouldn}.
To the best of our knowledge, while research has identified signals used by dependents, less work has focused on proposing measures developers can use. 
Existing work has focused on determining package signals, such as packages in decline~\cite{mujahid2021toward} or deemed unhealthy~\cite{linaaker2022characterize}.
However, signals designed to help developers prioritize dependency review efforts in changes are also needed to provide more immediate feedback to dependents and to aid in selecting dependencies for review.
Building upon prior work, we analyze whether contributor reputation~\cite{bosu2014impact}, proxied with network centrality measures, can serve as a signal for developer review of upstream changes in dependencies.

\subsection{Developer Social Networks}\label{bac:dsn}

Social networks can be created with individuals as vertices and their relations as edges and analyzed through graph theory~\cite{scott2012social}.
In a developer social network, the vertices are actors in the software development process (e.g., developers), and the edges are connections among the actors (e.g., collaboration or communication)~\cite{herbold2021systematic}. 

A rich body of literature has studied developer social networks~\cite{bosu2014impact, herbold2021systematic, meneely2011socio}.
Developer social networks have also been studied within GitHub.
Graphs have been defined with contributors as vertices and contributor collaboration or interactions as edges.
Edges were created if developers worked on the same projects~\cite{thung2013network, casalnuovo2015developer}, collaborated through pull requests~\cite{el2019empirical, sapkota2019network}, and followed each other~\cite{blincoe2015ecosystems}.
On the other hand, networks have been constructed for software packages, with packages as vertices and dependencies as edges~\cite{kikas2017structure, decan2019empirical, zimmermann2019small}.
As such, we built upon prior work constructing a contributor collaboration network in GitHub projects that host the top Rust packages with contributors as vertices and contributor collaboration as edges.
 
Social network measures can be calculated.
Among them, centrality measures can quantify what vertices are central within the graph~\cite{scott2012social}.
Centrality measures have been used in various fields to indicate the central actors in the network, including but not limited to networks of researchers, criminals, and students~\cite{das2018study}. 
Centrality measures are a common approach to proxy constructs such as reputation, trust, and influence in prior research, including in software engineering~\cite{bosu2014impact, meneely2011socio, asim2019trust, ceolin2017social, meo2017using, zahi2020improved, csimcsek2020combined}.
Hence, we leverage centrality measures as a proxy for contributor reputation.

%% file: sections/3.0.methods.tex
\section{Methodology}
\label{sec:method}

In this section, we present our research methodology that leverages mixed methods.
First, we constructed a developer network and calculated network centrality measures (Section~\ref{sec:network}).
Then, we conducted a survey questionnaire to understand how developers choose to review dependencies (Section~\ref{sec:questionnaire}).
Finally, we model network centrality measures to predict developers' review of dependencies (Section~\ref{sec:model}).

%% file: sections/3.1.network.tex
\subsection{Rust Developer Network}
\label{sec:network}

The following steps were performed to construct our developer network: ecosystem selection, data collection, developer social network construction, network properties exploration, and centrality measures calculation.

\subsubsection{Ecosystem Selection}

The Rust ecosystem has an active culture of collaboration and use of open source packages~\cite{schueller2022evolving}.
Many security-critical projects, like popular blockchain networks, are being developed in Rust~\cite{yakovenko2018solana, wood2016polkadot}. 
Further, the Rust ecosystem has built multiple tooling frameworks to help the secure usage of packages~\cite{rustsecure}. 
Therefore, we chose to study the Rust ecosystem in this paper.

\begin{table}
    \centering
    \caption{Overview of the collected data 
    to construct a developer network in Rust.}
    \input{tables/collected_data}
    \label{tab:dataset}
\end{table}

\subsubsection{Data Collection}

We obtained the metadata for packages from Crates.io from the official data dump~\cite{cratesio}. 
At the time of data collection of the study, from October 2022 to March 2023, Crates.io hosted 92,231 packages. 
We chose the most 1,000 downloaded packages and all their dependency packages, which resulted in 1,724 packages. 
While ideally, the developer network should include collaboration history from all the Crates.io packages and all open source projects collaborated by the package developers; we chose to work on a subset of the packages as our data collection methodology is constrained by GitHub REST API~\cite{githubrest} rate limit. 
We discuss the limitations 
in Section~\ref{limitation}.
The top packages were chosen to find relevant packages for the Rust community.

Out of the 1,724 packages, we found valid GitHub repositories for 1,644 packages using the \textit{package-locator} tool~\cite{packagelocator2024}.
We restricted our study to only GitHub repositories, as we identified distinct developers through GitHub accounts and collected code review information from GitHub. 
These 1,644 packages are hosted on 1,088 distinct repositories. 
Multiple packages may be stored in a repository, thus leading to more packages than repositories being found.
To build the developer network, we considered code activities over two years, between October 2020 and October 2022. 
Specifically, we collected 109,512 commits from this period. 
We selected two years for the data collection, following prior work~\cite{meneely2011socio}, as social network analysis is more expensive with more data.

For each commit, we collected the author's GitHub user account. 
We consider each distinct GitHub user as an individual developer, excluding the bot accounts~\footnote{The bot accounts' usernames are suffixed by `[bot]' on GitHub.\label{foot:bot}}. 
We also determined if each commit was reviewed by developers other than the author. 
We consider a commit to be code reviewed if there is a review approval on the associated pull request on GitHub or the commit was merged into the codebase by a different developer~\cite{imtiaz2023Your}. 
We identified 53\% (57,592) of the studied commits to have been code reviewed. 
We also collected data on the rejected pull requests, as the accepted pull request covers the previously collected commits.
If a non-merged pull request has a review from or was closed by a different developer, we consider that as a collaboration between the author and reviewer. 
In total, we obtained 2,975 pull requests rejected by a reviewer.

Overall, we found 6,949 distinct developers in our data set who are authors or reviewers of the studied commits.
Of these, 6,616  developers have authored at least one commit. 
Similarly, 2,891 developers have reviewed a commit at least once. 
We construct a social network over these 6,949 developers in the following section.
Table~\ref{tab:dataset} provides an overview of the dataset.

\subsubsection{Developer Social Network Construction}
\label{sec:network-construction}

Our developer social network is a graph data structure where vertices represent developer contributors and edges represent collaboration among contributors~\cite{herbold2021systematic}. 
We focus on this study on contributor collaboration with the rationale that collaboration can be mined directly from the repository history, unlike communication that can spread across various channels, such as emails and Slack. 
We use two metrics to capture contributor collaborations~\cite{meneely2011socio, kerzazi2016can}.

\begin{enumerate}
    \item \textbf{File Co-edition Collaboration:} If two contributors work on the same file within a 30-day window, we consider that activity to indicate collaboration.
    We create a collaboration edge between every contributor that collaborated during the window.
    This metric was validated by Meneely et al.~\cite{meneely2011socio}.

    \item \textbf{Author-Reviewer Collaboration:}  If a contributor has reviewed code changes submitted by another, we consider that activity to indicate collaboration between the author and the reviewer.  
\end{enumerate}

We identified 166,675 times where two contributors had made changes to the same file within 30 days, involving a relationship between 18,461 contributor pairs. 
We also identified 62,320 cases where a contributor had reviewed code from another contributor, involving 14,363 relationships. 
Combining both types of relationships (the same contributor pair can have both types of relationships), we have edges between distinct 26,448 contributor pairs. Further, we set edge weights in the network as the number of collaborations following prior work~\cite{kerzazi2016can, meneely2011socio}. Finally, we constructed an undirected, weighted graph based on our collected data using the Python package \textit{networkx}~\cite{hagberg2020networkx}.

\begin{table}
    \centering
    \caption{Rust developer network structural overview.}
    \input{tables/network_overview}
    \label{tab:dsn}
\end{table}

\begin{figure}
    \centering
    \includegraphics[scale =0.4]{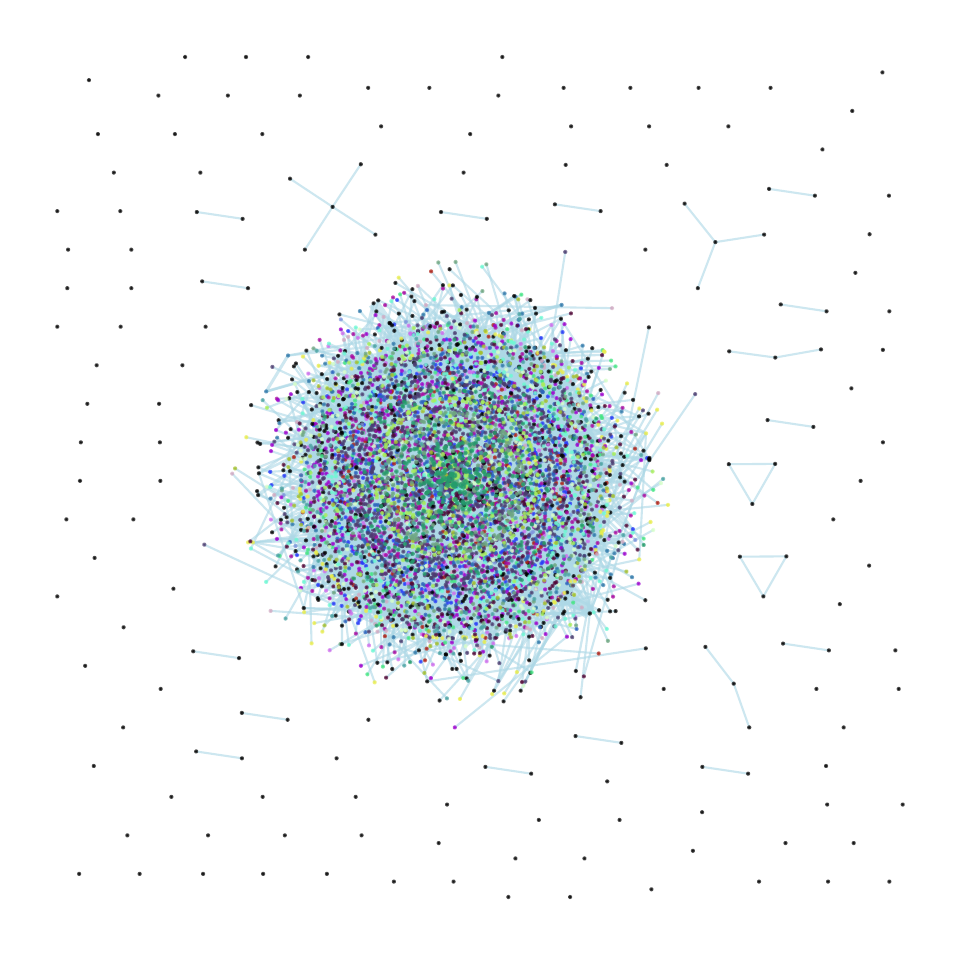}
    \caption{Developer network of Rust.}
    \label{fig:dsn}
\end{figure}

\begin{table*}[ht]
    \centering
    \caption{Centrality measures used in this study.}
    \input{tables/centrality_measures}
    \label{tab:centralitymeasures}
\end{table*}

\subsubsection{Network Properties Exploration}

The constructed network comprises 6,949 vertices (contributors) and 26,448 edges (collaborations). 
The largest connected component~\footnote{A connected component in a graph is a group of vertices where every two vertices have a direct or indirect path (via other vertices) between them.} in the network consists of 6,789 vertices (97.7\%), showing the majority of the studied contributors are interconnected. 
The rest of the contributors are isolated or connected by a small group of collaborators. 
The density~\footnote{The ratio of existing edges to all possible edges.} of our network is 0.001. 
The average clustering coefficient~\footnote{The clustering coefficient of a vertice is a measure of the extent to which its neighbors are also connected.} of the vertices is 0.4, typical of real-world social networks~\cite{walsh1999search}. 
The low density and clustering coefficient values indicate the graph consists of many tightly-knit clusters. 

The average shortest path length in the largest connected component is 3.7, meaning any two contributors are connected via approximately 4 other contributors on average, indicating a small-world phenomenon~\cite{walsh1999search, sherchan2013survey}. 
Further, we found 164 communities~\footnote{A subset of vertices within the graph such that connections between the vertices are denser than connections with the rest of the network.} using the Louvain community detection algorithm~\cite{que2015scalable}. 
However, most developers are members of 21 communities with at least 100 contributors. 
The community structure further establishes the interconnected nature of the Rust developer network. 
Table~\ref{tab:dsn} summarizes the properties of the network. 
Figure~\ref{fig:dsn} visualizes the network, where the vertices in the communities with larger than 100 contributors are assigned distinct colors. 
The network shows characteristics of a small world, which establishes the ground for centrality analysis~\cite{watts1998collective, wasserman1994social}, to be used in the sampling approach in Section~\ref{sec:survey-par}. 

\subsubsection{Centrality Measures Calculation}
\label{sec:centrality}

Centrality measures are a common approach to serve as a proxy for constructs such as reputation, trust, and influence (Section~\ref{bac:dsn}).
Overall, centrality measures indicate and rank the central actors within a network~\cite{riquelme2018centrality, cadini2009using}. 
In developer social network literature, Bosu et al.~\cite{bosu2014impact} studied the impact of developer reputation on code review outcomes by identifying the core developers in a network through six centrality measures. 
We incorporate five of those six measures in our work, as these were also used in a trust model proposed by Asim et al.~\cite{asim2019trust}. 
One metric we excluded from Bosu et al.'s~\cite{bosu2014impact} work is eccentricity, which only works for graphs with a single connected component and is unsuitable in our context. 
Table~\ref{tab:centralitymeasures} explains the five measures and provides the interpretation for each measure. 
The interpretation was inspired by Bosu et al.~\cite{bosu2014impact}'s work.

We compute the five measures using \textit{networkx} Python package that considers the weights of the edges. 
Thus, we get a normalized value between 0 and 1, representing the centrality of the contributor within the community according to each measure.
A higher value indicates a contributor that is more central to the Rust community network.
We then aggregate the measures for our survey to avoid bias from any single measure~\cite{csimcsek2020combined, bosu2014impact}. 
We take the normalized value of each of the five centrality measures listed in Table~\ref{tab:centralitymeasures}, then calculate the normalized sum of those values to get a final rating between 0 and 1.
Our approach of a simple additive weighting of normalized scores is commonly followed when combining multiple metrics for ranking~\cite{tofallis2014add, shabrina2022investigating, cody2018investigation}.

%% file: tables/collected_data.tex
\begin{tabular}{lr}
    \toprule
    Crates.io & 1,644 \\
    Repositories & 1,088 \\
    Commits & 109,512 \\
    Reviewed Commits & 57,592 (52.6\%) \\
    Rejected Pull Requests & 2,975 \\ 
    \midrule
    Developers & 6,949 \\
    Authors & 6,616 (95.2\%) \\
    Reviewers & 2,891 (41.6\%) \\
    \midrule
    Relationships & 26,448 \\
    File Co-edition Relationships  & 18,461 (69.8\%)\\
    Author-Reviewer Relationships & 14,363 (54.3\%) \\
    \bottomrule
\end{tabular}

%% file: tables/network_overview.tex
    \begin{tabular}{lr}
    \toprule
        Vertices & 6,949 \\
        Edges & 26,448 \\
        Components & 132 \\
        Isolates & 111 \\
        Network density & 0.001 \\
        Avg. clustering coefficient & 0.398\\
        Total communities & 164 \\
        \makecell[l]{Communities with more than a hundred vertices} & 21 \\
        \midrule
        \multicolumn{2}{c}{\textit{Largest Component}} \\
        \midrule
         Vertices & 6,789 (97.7\%) \\
         Edges &  26,417 \\
         Avg. shortest path length & 3.674 \\
    \bottomrule
    \end{tabular}

%% file: tables/centrality_measures.tex
    \begin{tabular}{p{3.5cm}p{5.5cm}p{7.5cm}}
    \toprule
        \textbf{Metric} & \textbf{Definition} & \textbf{Interpretation}  \\
        \midrule
        Degree centrality~\cite{freeman1977Set} &  The fraction of the total vertices in the network a vertice is connected to. & The extent of direct collaboration with other contributors indicates a contributor's number of direct interactions. \\
        \midrule
        Closeness centrality~\cite{freeman1979centrality} & The reciprocal of the average shortest path distance from the vertice to all other reachable vertices in the network. & The reciprocal average number of collaboration relationships needed to reach other contributors indicates how quickly a contributor can reach the entire community.\\
        \midrule
        Betweenness centrality~\cite{brandes2001Faster}  & The fraction of all pairs of vertices in the network whose shortest paths pass through that vertice.
        & The number of contributors that pass through the contributor as their shortest collaboration path to others indicates how often a developer acts as an intermediary. \\
        \midrule
        Eigenvector centrality~\cite{bonacich2007Some} & The centrality of a vertice is calculated based on the centrality of its neighbors.  & The direct collaboration with higher-ranked contributors in the network indicates contributor influence.\\ 
        \midrule
        PageRank~\cite{brin1998Anatomy} & A variation of eigenvector centrality, using the number and quality of edges. & The contributors who engage in higher-ranked collaborations within the network indicate contributor influence. \\  
        \bottomrule
    \end{tabular}

%% file: sections/3.2.survey.tex
\subsection{Survey Questionnaire}\label{sec:questionnaire}

To survey developers through a questionnaire, we had the following steps: questionnaire design, participant and sample selection, and response analysis. 

\begin{table*}[!htp]
    \centering
    \caption{The survey with the questions and options for answers.}
    \input{tables/survey}
    \label{tab:survey_questions}
\end{table*}

\subsubsection{Questionnaire Design}

The survey was designed iteratively between two researchers to ask objective questions developers could answer. 
We ask how developers, as downstream dependents, choose to review dependencies.
Our survey contained three main closed-ended questions (SC). 
We also provide options for respondents to provide additional comments in open-ended questions (SO).
Table~\ref{tab:survey_questions} shows the survey questions and corresponding Likert-scale options. 
The motivation behind the closed-ended questions is the following:

    \indent \textbf{SC1.} 
    The motivation behind SC1 is first to understand what general strategies developers choose to review upstream changes before integrating them into their codebase. 
    We also want to know if contributors' identities impact their review process. 
    For example, we ask if contributors' reputations within the Rust community and the respondents' familiarity with contributors' work impacts the developers' review process. 

    \indent  \textbf{SC2.} 
    The motivation behind SC2 is to 
    understand how developers recognize and know contributors within the Rust community network. 
    To this end, we independently sample ten contributors from the network for each recipient and ask them how well they know the contributors. 
    We provide five Likert-scale options for the recipients to indicate how they know the contributors in the network.

    \indent  \textbf{SC3.} The motivation behind SC3 is to understand how developers review upstream changes based on contributor identity. 
    We present the survey recipient with the same ten contributors from SC2 and ask how the recipient would review dependency changes from them. 
    We provide four Likert-scale options to indicate the level of review the recipient would put into upstream code changes from each contributor.
    SC3 answers complement the general strategy answers of SC1 by gathering how developers would choose to review changes from specific contributors.

\subsubsection{Participant and Sample Selection}\label{sec:survey-par}

We sent the survey to a subset of 1,995 developers within the network created in Section~\ref{sec:network-construction}.
We selected developers who have directly collaborated with at least 5 others in the network with a valid email address.
We subset the sample with the rationale that developers who only had a few interactions in the community may be unable to provide informed answers about others.
Contributors in the network collaborated with a median of 2 contributors and an average of 7.61. Thus, we select 5 as an intermediate value between both as our threshold.
We received 285 responses, with a response rate of 14.3\%. 

Additionally, we independently sample ten contributors from the network for each respondent to ask (i) how well they know these contributors for SC2; and (ii) how the recipient would review upstream code changes from them for SC3. 
To this end, we sample ten contributors for each respondent in the following ways: (i) five contributors from the direct collaborators of the recipient as indicated by an edge in our constructed network; and (ii) five contributors from the rest of the network with whom the recipient does not have an edge with.
Within these two groups (direct collaborators and others), we sample 3 from the top 50 according to the aggregated centrality measure described in Section~\ref{sec:centrality}, and two from the rest of that group. 
We chose a purposeful sampling approach~\cite{baltes2022sampling} so the recipients had a higher chance of being familiar with at least one of the contributors. 
Further, this approach ensures that the recipients have top, average, and low-ranked contributors, based on the centrality score, to provide their opinions.

\subsubsection{Response Analysis}

We quantitatively and qualitatively analyze our survey answers.
We aggregate responses and apply standard statistical tests for our quantitative closed-ended answers (SC1, SC2, SC3).
Meanwhile, we start our qualitative analysis with the first and third authors collaboratively creating inductive codes for the open-ended questions (SO1, SO2, SO3), creating an initial codebook.
We coded two aspects from the answers: 
(i) factors other than contributor reputation considered when reviewing dependencies, and 
(ii) reasons developers choose to review their dependencies or not.
The first aspect is coded from question SO1, while the second is from questions SO2 and SO3.
Then, both authors initially individually coded all survey responses.

We then followed an iterative process until our threshold of the inter-coder agreement was exceeded using Krippendorff's alpha~\cite{krippendorff2018content}.
In each phase, disagreements were resolved through discussion, resulting in a refined coding procedure and codebook.
The authors would then recode the survey responses based on the discussions and recalculate the inter-coder agreement.
Four iterations were held between authors until we achieved an alpha of 0.82.
The alpha value exceeded the threshold of 0.8.
Based on the threshold, we can consider that there is statistical evidence of the evaluation reliability~\cite{gonzalez2023reliability}.
Lastly, the first author re-coded the remaining disagreements.
We present relevant codes to help contextualize the survey responses.
The generated codebook is available in the supplemental material~\cite{us2024Supplemental}.
We present the survey results in Section~\ref{sec:rq1}.

%% file: tables/survey.tex
\begin{tabularx}{\linewidth}{lBS}
    \toprule
    \textbf{ID} & \textbf{Question} & \textbf{Options}\\
    \midrule
    \multirow{4}{*}{\textbf{SC1}}  & Please rate the below statements based on how you review the incoming upstream code changes in your dependency packages (when adding or updating a package). --- The question presents six statements which are listed in Table~\ref{tab:q5} when discussing the findings. & \multirow{4}{=}{Always, Often, Sometimes, Rarely, Never}\\
    \midrule
    \multirow{2}{*}{\textbf{SO1}} & What other factors impact how carefully you review the upstream code changes before adding/updating a package? & \multirow{2}{=}{\textit{Open-ended}}\\
    \midrule
    \multirow{8}{*}{\textbf{SC2}} & \multirow[b]{3}{=}{In the Rust package ecosystem context, choose one of the five options for the below-listed GitHub users. } & 1. I have never heard of this person before. \\
    &  & 2. I recognize this name, but I don't know much about them. \\
    & \multirow{3}{=}{--- The respondent is presented with a list of ten GitHub users who are sampled separately for each respondent.} & 3. I know this person, but I don't know anyone who has worked with them.\\
    & \multirow{3}{=} {The sampling strategy is explained in Section~\ref{sec:survey-par}.} & 4. I know this person, and I have worked with people who have worked with them. \\
    &  & 5. I have directly worked with this person.\\
    \midrule
    \multirow{8}{*}{\textbf{SC3}} & \multirow[c]{2}{=}{When adding or updating a package, how carefully would you review upstream code changes coming from the below-listed GitHub users? 
    } &  1. I would not include code from this person in my project. \\
    & \multirow[c]{5}{=}{Please choose one of the four options based on your knowledge/association with them. Note that the code changes are within the dependency packages of your project. For example, when updating a package, how do you review the new changes that come with the update?} & 2. I would review each line of upstream code change coming from this person. \\
    & & 3. I would skim through the upstream code changes coming from this person. \\
    & \multirow[b]{2}{=}{--- The respondent is presented with the same list of ten GitHub users from SC2.} & 4. I do not feel the need to review the upstream code changes coming from this person.\\
    \midrule
    \multirow{2}{*}{\textbf{SO2}} & Can you explain your reasoning if you have chosen the same option for every user? Skip if not applicable. & \multirow{2}{=}{\textit{Open-ended}}\\
    \midrule
    \textbf{SO3} & Please add anything more you have to say. & \textit{Open-ended}\\
    \bottomrule
\end{tabularx}

%% file: sections/3.3.model.tex
\begin{table*}[!bt]
    \centering
    \caption{Likert scale responses on how strategies impact review process of upstream code for SC1.}
    \input{tables/SC1}
    \label{tab:q5}
\end{table*}

\subsection{Regression Modeling}
\label{sec:model}

We determine the influence of the network centrality measures as a signal for how developers choose to review upstream changes based on contributor identity through regression modeling.
We select the respondents who, in SC3, answered for all the presented GitHub users (shown in Table~\ref{tab:survey_questions}).
In our analysis, we control for the fact that multiple responses for the same contributor can come from different respondents. 
To this end, we run a mixed effect regression analysis~\cite{faraway2016extending} on the responses.
Mixed-effect linear regression models the association between fixed effect variables and dependent variables when clusters or categorizations are present in the data. 
The regression model eliminates the impact of the clusters called random effects and represents only the correlation between the fixed effect and the dependent variables.

We create a multivariate mixed-effect linear regression, adding the contributor identity as a random effect variable in the regression analysis since multiple responses for the same contributor can introduce clusters in the data.
We use the R package \textit{lme4}.
In our model, we define the level of the review indicated by the respondents for a given contributor as the dependent variable and the centrality measures of the contributors as the fixed effect variables.
Fixed effects with a variance inflation factor (VIF) were excluded to account for multicollinearity in the models.
Thus, we excluded the degree centrality measure due to exceeding our VIF threshold of five.
To reduce heteroscedasticity, we checked if any fixed-effect variable had a high variation requiring a logarithmic transformation.
We did not apply any transformation, as no variable had a variance over 0.2 (centrality measures were already normalized and ranged between 0 and 1).
To assess the goodness of fit of the models, we use the mixed effect specific variants of $R^2$ proposed by Nakagawa and Schielzeth~\cite{nakagawa2013general}, marginal ($R^2_m$) and conditional ($R^2_c$).
$R^2_m$ represents the explainable variance by the fixed factors, while $R^2_c$ is the explainable variance by the fixed and random factors.
We report the number of observations ($O$) and groups ($G$) for the random effects.
We also report the significant coefficients for the variables with p-values using ANOVA and the sum of squares.
The regression modeling results are presented in Section~\ref{sec:rq2}.

%% file: tables/SC1.tex
\begin{tabular}{lp{0.5\linewidth}p{0.3\linewidth}}
\toprule
     \textbf{ID} & \textbf{Statement}  &  \textbf{SC1 Distribution}   \\
\toprule
SC1-1 & I review the incoming upstream code changes before adding/updating a package.        & 
  \vspace{-2.75pt} \distributionsurvey{0.08}{0.16}{0.36}{0.30}{0.11}{0}{24\%}{40\%} \\
SC1-2 &
  I look at the authors and reviewers~\tablefootnote{In the survey, we specifically ask about changes' authors and reviewers. In the paper, we refer to authors and reviewers as contributors\label{foot:cont}.} of the incoming upstream code changes before adding/updating a package &
  \vspace{-2.75pt} \distributionsurvey{0.11}{0.17}{0.23}{0.27}{0.22}{0}{29\%}{49\%} \\
SC1-3 &
  The authors'/reviewers'~\footref{foot:cont} reputations within the Rust ecosystem impact how carefully I review before adding/updating a package. &
  \vspace{-2.75pt} \distributionsurvey{0.255}{0.255}{0.23}{0.11}{0.15}{0}{51\%}{26\%} \\
SC1-4 &
  My past collaborations with the authors/reviewers~\footref{foot:cont} impact how carefully I review before adding/updating a package. &
  \vspace{-2.75pt} \distributionsurvey{0.13}{0.32}{0.20}{0.13}{0.22}{0}{45\%}{35\%} \\
SC1-5 &
  My familiarity with the authors'/reviewers'~\footref{foot:cont} past work impacts how carefully I review before adding/updating a package. &
  \vspace{-2.75pt} \distributionsurvey{0.22}{0.28}{0.22}{0.12}{0.16}{0}{50\%}{28\%} \\
SC1-6 & I do not review the incoming upstream code changes before adding/updating a package. & 
  \vspace{-2.75pt} \distributionsurvey{0.12}{0.34}{0.21}{0.16}{0.17}{0}{46\%}{33\%} \\
\midrule
\multicolumn{3}{c}{\legendtable{Always}{blue2} \legendtable{Often}{blue1} \legendtable{Sometimes}{gray1}\legendtable{Rarely}{pink1} \legendtable{Never}{pink2}} \\
\bottomrule
\end{tabular}

%% file: sections/4.RQ1.tex
\section{RQ1 Developer Dependency Review}
\label{sec:rq1}

\textbf{How do developers choose to review dependencies in the Rust Ecosystem? }
We investigate RQ1 through responses from our developer survey presented in Section~\ref{sec:questionnaire}, consisting of three closed-ended questions in Table~\ref{tab:survey_questions}. 
The questions help us to investigate: 
(i) what dependency review strategies are chosen by developers; 
(ii) whether developers recognize collaboration activity in the constructed Rust community network;
and (iii) how contributor identity impacts developers' level of review of upstream code.
Additionally, the survey included three open-ended questions that complemented the responses to the closed-ended questions.
We present the findings in this section.

\subsection{Dependency Review Strategies}\label{sec:qa}

In SC1, we investigate developers' strategies when reviewing upstream code when adding and updating a dependency through a Likert-scale matrix of six statements. 
Table~\ref{tab:q5} lists the six statements and the response distribution. 

Only 24\% of the respondents answered that they always or often review the upstream changes.
Hence, respondents are unlikely to review the upstream code changes before integrating them into their projects. 
The respondents provided reasons in the open-ended answers for SO2 and SO3.
From the 180 answers, we found 17 reasons. 
The most mentioned are shown in Table~\ref{tab:reasons}.
Among the reasons developers infrequently or do not review dependencies is that reviewing all changes in a dependency is hard (29 responses).
Projects have many dependencies, even more so when you consider transitive dependencies.
As such, reviewing all changes requires considerable time, company support, and tooling to streamline the process.
Respondents even mentioned that as contributions to open-source are voluntary, there should be no expectation for such work without a support contract (9 responses).
The rarity of vulnerabilities or malicious code in dependencies is considered not worth the effort of reviewing upstream (5 responses).
In other cases, reviews are not needed due to the context of the dependent or the dependency (32 responses).
For example, when developing toy projects, reviewing dependencies is not needed.
Additionally, a higher level of review is applied when reviewing projects developers contribute to rather than dependencies.
Similarly, only 29\% of the respondents answered that they always or often look at the contributors of the upstream changes before adding or updating a package. 

 \begin{table*}[tb]
    \centering
    \caption{Top seven reasons developers review or not dependencies mentioned in SO2 and SO3 with their SC1 answers.}
    \input{tables/top_reasons}
    \label{tab:reasons}
\end{table*}

\begin{table}[t]
    \centering
    \caption{Top 10 non-contributor reputation dependency review factors mentioned in SO1.}
    \input{tables/top_factors}
    \label{tab:package-signals}
\end{table}

While the survey respondents are less likely to review all upstream code changes, developers indicated that the contributors' reputations in the Rust community impact their decision process before adding or updating a package.
\textbf{More than half of the respondents (51\%) suggested that the reputations of the upstream contributors impact always or often how carefully they review upstream changes.} 
In contrast, only 26\% of the respondents indicated that the contributors' reputation rarely or never impacts their review process. 
Similarly, respondents indicated that familiarity with the upstream contributors' work (50\%) and past collaborations (46\%) impact their review process.

Respondents also mentioned reasons why some level of review is employed.
Rather than developers thoroughly reviewing the code during updates, reviews skim through the changes to determine if anything noteworthy occurred (44 responses).
For example, through  \textit{changelogs}~\footnote{A natural text documentation of the changes in a new version.}.
Emphasis is placed on reviewing projects when adding a dependency compared to updates (28 responses).
Factors, including contributor reputation, determine which dependency should be adopted.
Developers also mentioned other non-contributor reputation factors in SO1.
We received 185 responses, finding 28 factors.
The most mentioned are shown in Table~\ref{tab:package-signals}.
Respondents thus trust the ecosystem's projects, community, and tools to provide security guarantees to dependents (32 responses).
Some respondents mentioned that ecosystem trust may be misplaced, expecting a robust safety net.

Respondents who indicated that they never looked at incoming upstream changes, mostly (72\%), did not look at contributors.
Additionally, most also did not review the upstream based on contributors' reputation (68\%), past collaboration (64\%), and familiarity (60\%).

\subsection{Recognition of Collaborators}\label{sec:qb}

The objective behind SC2 was to investigate how developers recognize and know the contributors with whom they collaborated in the network.
Considering ten responses for the ten given developers from each respondent in SC2, we have 2,844 total data points. 
In 1,375 cases, our constructed network has an edge between the respondent and the given contributor. 
However, only in 244 cases (18\%) the respondent answered that they directly worked with the given contributor. 
In another 566 cases (41\%), the respondent indicated some familiarity with the given contributors (any option other than \textit{I have never heard of this person before}). 
In the remaining 565 cases (41\%), the respondents stated they had never heard of the person before.

While the file co-edition relationship may not ensure the two developers are, indeed, familiar with each other, we find that the survey respondents failed to recognize contributors in the case of author-reviewer relationships as well. 
Out of the 1,375 cases, there are 736 cases where the network shows an author-reviewer relationship between the recipient and the given contributor. 
However, only 214 (29\%) of these cases indicated a direct collaboration, while the respondents showed some familiarity in another 435 (59\%) cases. 
In the remaining 87 cases (12\%), the respondents stated they had never heard of the person before, even though our data shows an author-reviewer relationship between them. 
Conversely, there are 1,469 cases where the network does not show an edge between the survey respondent and the given contributor. 
In only 32 cases (2\%), however, the respondents indicated they had worked with the contributor.
Given the distributed nature of open-source development, a developer may not be familiar with all others working on the same project simultaneously. 
Further, a developer may not remember every collaborator. 

\begin{figure}[t]
    \centering
    \includegraphics[scale=0.48]{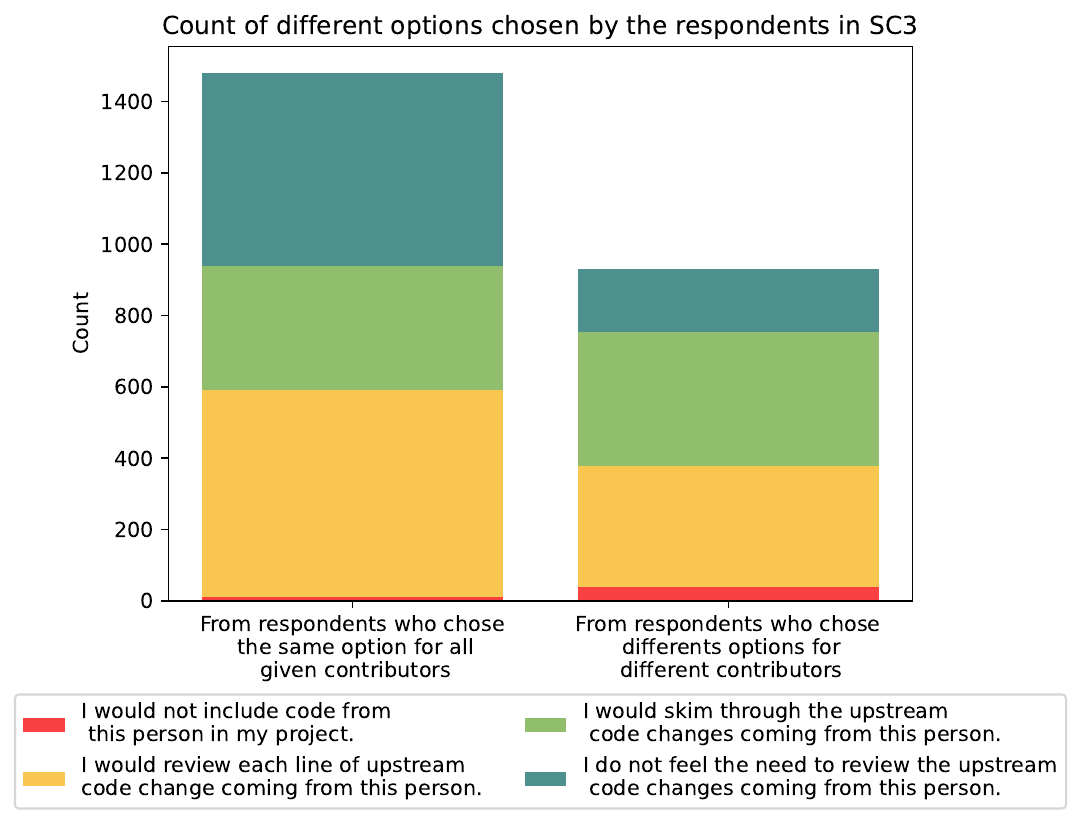}
    \caption{Count of options chosen by the respondents in SC3.}
    \label{fig:q4}
\end{figure}

\subsection{Review Based on Contributor Identity}\label{sec:qc}

In SC3, we are interested in how respondents choose to review upstream code using contributor identity.
Of the 285 respondents, 44 were incomplete and did not fill out an option for all given contributors.
Most incomplete responses did not select an option for any contributor (33 respondents).
In the other cases, nine respondents selected only one option and two respondents filled out at least seven out of ten options.
We present our analysis for the remaining 241 complete responses.

We find that 148 respondents (61\%) chose the same option for all the contributors, corresponding to 1,480 cases.
On the other hand, 93 respondents chose different options for different contributors, indicating that the contributor's identity may impact the level of review they employ for upstream changes.
In total, 930 cases were selected for those who considered contributor identity.

Figure~\ref{fig:q4} shows the response distribution from both types of respondents. 
We conducted a Chi-squared test to determine if there were differences in the chosen level of the review between respondents who chose the same option for all contributors and respondents who chose different options for different contributors.
The distribution of responses was statistically significantly different between both groups ($ X^2 = 147.85, df = 3, p < 0.001$).
Notably, 540 (37\%) cases that chose the same option indicated not feeling the need to review upstream code from the person, compared to the 177 (19\%) cases that chose different options.
Additionally, 350 (24\%) cases that chose the same option would skim through the changes made by the person, compared to the 376 (40\%) cases that chose different options.
\textbf{Hence, although respondents who did not consider contributor identity were more likely not to review upstream code, respondents who did consider identity were more likely to review by skimming upstream changes.}

 \begin{table*}[tb]
    \centering
    \caption{The effect of contributor reputation on the level of review of dependencies.}
    \input{tables/models}
    \label{tab:melr}
\end{table*}

\begin{figure*}[tb]
     \centering
     \begin{subfigure}[b]{0.32\textwidth}
         \centering
         \includegraphics[width=\textwidth]{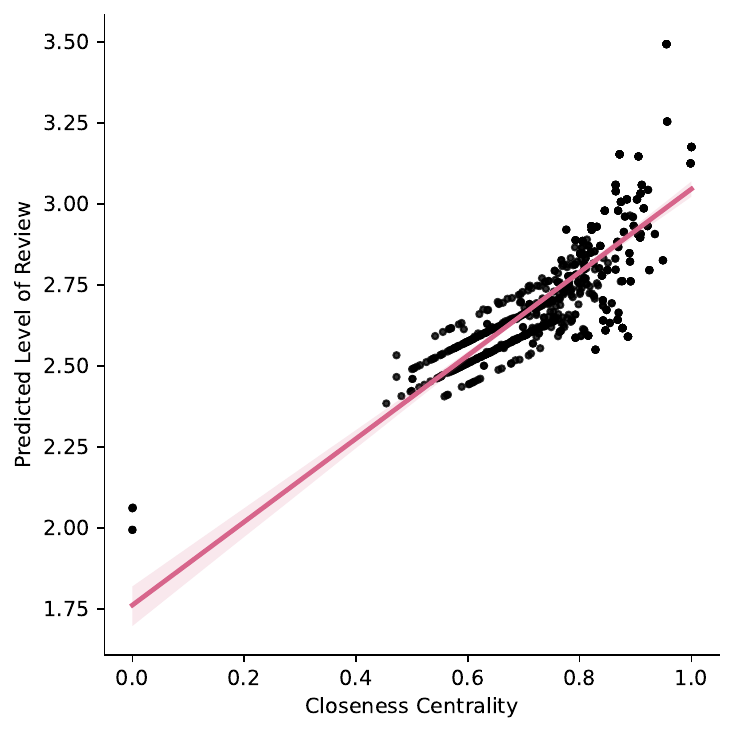}
         \caption{Different level of review}
         \label{SUBFIG:ModelDifferent}
     \end{subfigure}
     \hfill
     \begin{subfigure}[b]{0.32\textwidth}
         \centering
         \includegraphics[width=\textwidth]{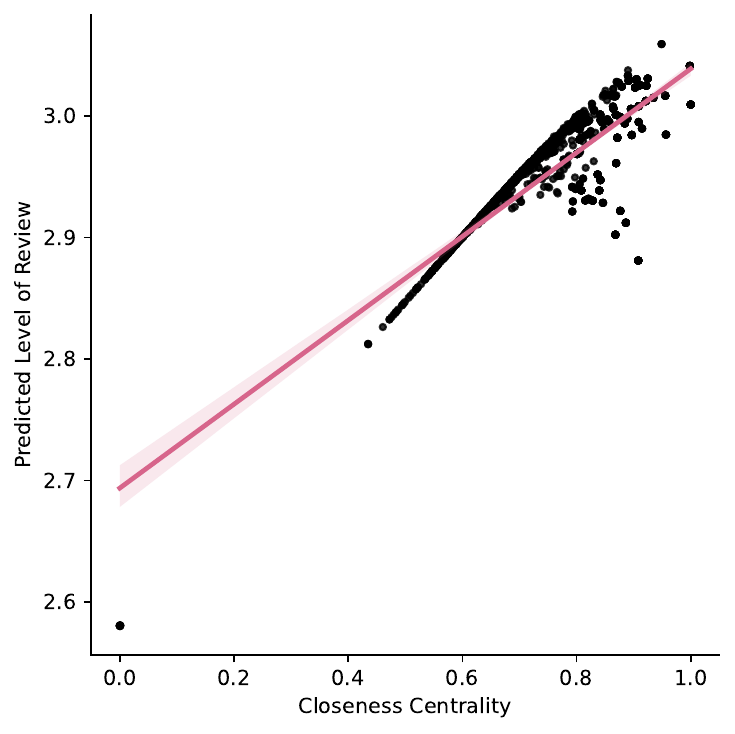}
         \caption{Same level of review}
         \label{SUBFIG:ModelSame}
     \end{subfigure}
     \hfill
     \begin{subfigure}[b]{0.32\textwidth}
         \centering
         \includegraphics[width=\textwidth]{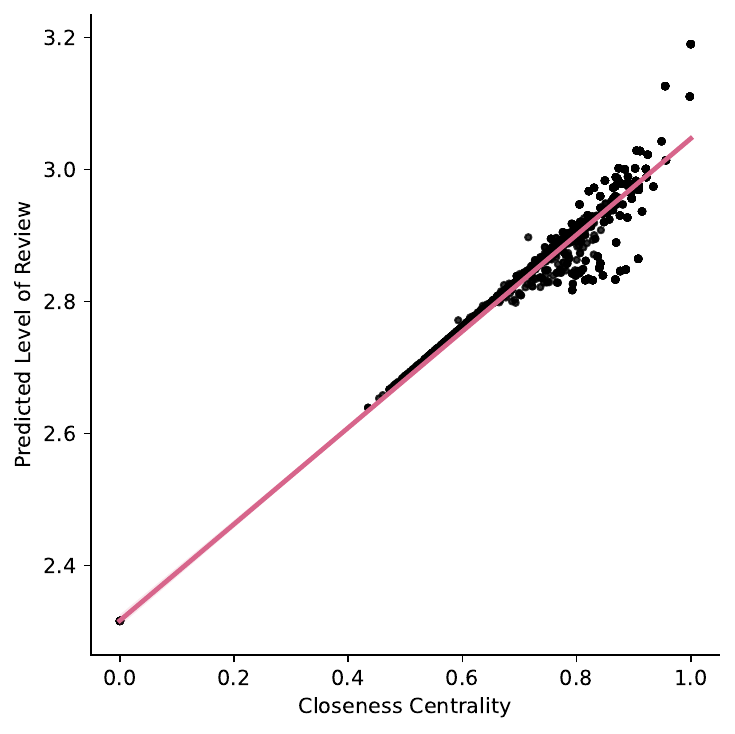}
         \caption{All}
         \label{SUBFIG:ModelAll}
     \end{subfigure}
    \caption{Predicted level of review of the dependency relationship with closeness centrality.}
    \label{fig:model}
\end{figure*}

Reasons were provided in the open-ended answers of SO2 and SO3.
Respondents appeared to misinterpret our questions and answered based on how they would review pull requests in their projects rather than changes in upstream code.
Consequently, the respondents did not discriminate based on identity by reviewing all lines in a change (36 responses).
Respondents indicated they would review all project changes, as anyone can make mistakes when contributing to a project (13 responses).
We did not discard the answers of the respondents with possible misinterpretations.
We discuss our reasoning in Section~\ref{limitation}.

%% file: tables/top_reasons.tex
\begin{tabular}{p{0.45\linewidth}p{0.24\linewidth}p{0.24\linewidth}}
\toprule
\textbf{Codes (\# Respondents) -- Summary description} &
  \textbf{SC1-1 Distribution} &
  \textbf{SC1-3 Distribution} \\
\toprule
\textit{Light skim updates} (44) -- While reviewing a dependency update, the developer lightly skims the changes (e.g., changelog). &
  \vspace{-3pt} \distributionsurvey{0.02}{0.15}{0.37}{0.41}{0.05}{0}{17\%}{46\%} &
  \vspace{-3pt} \distributionsurvey{0.17}{0.27}{0.24}{0.15}{0.17}{0}{44\%}{32\%} \\
\midrule
\textit{Review the same for everyone} (36) -- The developer treats every contributor the same during review.  &
  \vspace{-3pt} \distributionsurvey{0.21}{0.09}{0.47}{0.12}{0.12}{0}{30\%}{24\%} &
  \vspace{-3pt} \distributionsurvey{0.06}{0.35}{0.18}{0.09}{0.32}{0}{41\%}{41\%} \\
\midrule
\textit{Trust ecosystem} (32) -- The developers trust and depend on the ecosystem to provide security guarantees for dependencies. For example, dependency maintainers will review the dependencies. &
  \vspace{0.55pt} \distributionsurvey{0.03}{0.095}{0.345}{0.41}{0.12}{0}{12\%}{53\%} &
  \vspace{0.55pt} \distributionsurvey{0.06}{0.31}{0.16}{0.25}{0.22}{0}{37\%}{47\%} \\
\midrule
\textit{Project context-dependent} (32) -- The level of review of a project varies on the context of the project (e.g., toy project, security-critical dependency).  &
  \vspace{0.55pt} \distributionsurvey{0.00}{0.07}{0.27}{0.53}{0.13}{0}{ 7\%}{66\%} &
  \vspace{0.55pt} \distributionsurvey{0.133}{0.233}{0.20}{0.20}{0.233}{0}{36\%}{43\%} \\
\midrule
\textit{Hard to review dependencies} (29) -- The dependencies are hard to review for developers due to reasons including the number of dependencies, time available, and tools.  &
  \vspace{0.55pt}\distributionsurvey{0.07}{0.11}{0.18}{0.43}{0.21}{0}{18\%}{64\%} &
  \vspace{0.55pt}\distributionsurvey{0.14}{0.32}{0.14}{0.11}{0.29}{0}{46\%}{40\%} \\
\midrule
\textit{Review when adding a dependency} (28) -- Once a dependency is reviewed during adoption, a lower level of review occurs for updates. Factors considered are mentioned in SO1.  &
  \vspace{0.55pt}\distributionsurvey{0.04}{0.00}{0.44}{0.37}{0.15}{0}{ 4\%}{52\%} &
  \vspace{0.55pt}\distributionsurvey{0.19}{0.26}{0.15}{0.15}{0.26}{0}{45\%}{41\%} \\
\midrule
\textit{Decide review level based on reputation} (26) -- The level of review considers contributor reputation, with respondents mentioning more relevance of the maintainers or leads reputation.  &
  \vspace{0.55pt}\distributionsurvey{0.04}{0.22}{0.43}{0.22}{0.09}{0}{26\%}{31\%} &
  \vspace{0.55pt}\distributionsurvey{0.174}{0.304}{0.304}{0.043}{0.174}{0}{47\%}{21\%} \\
\midrule
\multicolumn{3}{c}{\legendtable{Always}{blue2} \legendtable{Often}{blue1} \legendtable{Sometimes}{gray1}\legendtable{Rarely}{pink1} \legendtable{Never}{pink2}} \\
\bottomrule
\end{tabular}

%% file: tables/top_factors.tex
\begin{tabular}{lr}
\toprule
\textbf{Factor} & \textbf{\# Respondents}\\
\midrule
 Popularity & 52 \\
 Usage & 27 \\
 Project context & 25 \\
 Size & 24 \\
 Security implication & 21 \\
 Code quality & 19 \\
 API changes & 18 \\
 Maintenance & 18 \\
 Changelog & 16 \\
 Community & 16 \\
 \bottomrule
\end{tabular}

%% file: tables/models.tex
\begin{tabular}{lr@{\hskip 1pt}r@{\hskip 0pt}rr@{\hskip 1pt}r@{\hskip 0pt}rr@{\hskip 1pt}r@{\hskip 0pt}r}
\toprule
    & \multicolumn{3}{c}{\textbf{Different level of review}} & \multicolumn{3}{c}{\textbf{Same level of review}}  & \multicolumn{3}{c}{\textbf{All responses}}             \\
    & \multicolumn{3}{c}{O = 929, G = 551} 
    & \multicolumn{3}{c}{O = 1477, G = 795} 
    & \multicolumn{3}{c}{O = 2406, G = 1170} \\
    & \multicolumn{3}{c}{$R^2_m$ = 6\%, $R^2_c$ = 13\%} & 
    \multicolumn{3}{c}{$R^2_m$ $<$ 0.01\%,  $R^2_c$ =$<$ 0.01\%} & 
    \multicolumn{3}{c}{$R^2_m$ = 2\%,  $R^2_c$ = 2\%} \\ 
    \cmidrule(lr){2-4} \cmidrule(lr){5-7} \cmidrule(lr){8-10}
    & Coeffs (Err.) & & Sum Sq. 
    & Coeffs (Err.) & & Sum Sq. 
    & Coeffs (Err.) & & Sum Sq. \\
\midrule
\textit{(Intercept)}   
    & 1.99 (0.20) & & 
    & 2.58 (0.18) & & 
    & 2.32 (0.13) & & \\
Closeness Centrality 
    & 0.92 (0.29) & ** & 5.92 
    & 0.53 (0.26) & * & 3.29 
    & 0.74 (0.19) & *** & 10.93\\
Betweenness Centrality 
    & 0.53 (0.34) & & 1.42 
    & 0.05 (0.26) & & 0.03 
    & 0.24 (0.19) & & 1.14 \\
EigenVector Centrality 
    & -0.16 (0.18) & & 0.44 
    & -0.15 (0.14) & & 0.96 
    & -0.17 (0.10) & & 2.01 \\
PageRank Centrality 
    & 0.15 (0.31) & & 0.14 
    & -0.07 (0.23) & & 0.06 
    & -0.01 (0.17) & & 0.00 \\\bottomrule
\end{tabular}

    \vspace{.5ex}
    {\scriptsize Statistical significance using ANOVA: *** $p <0.001$, ** $p <0.01$, * $p <0.05$}
    \vspace{-1ex}

%% file: sections/5.RQ2.tex
\section{RQ2 Network Centrality Signals}
\label{sec:rq2}

\textbf{How can network centrality measures, as proxies for contributor reputation, signal the need to review dependencies?} 
We investigate RQ2 through the mixed-effect linear regression models constructed in Section~\ref{sec:model}.
We determine the influence of the network centrality measures (Table~\ref{tab:centralitymeasures}) on the level of dependency review employed by developers based on contributor identity (answers for SC3).
A score of 1 indicates the highest level of review for a dependency, while a 4 is the lowest.
The corresponding Likert-scale options for each level of review are listed in Table~\ref{tab:survey_questions}.
We create three different models.
The first model is for the 93 respondents who choose different options for the contributors.
Meanwhile, the second model is for the 148 respondents who chose the same option for all contributors.
Finally, the third model is for all 241 respondents who completed SC3.
The performance of the models is shown in Table~\ref{tab:melr}.

Our first model distributed 929 responses over 551 clusters of responses.
We find that the closeness centrality is a statistically significant fixed effect explaining how developers choose to review upstream changes for respondents.
The centrality measure indicates how quickly a contributor can reach the entire Rust community.
\textbf{Combined with the positive coefficient, the upstream code changes of contributors with higher closeness centrality significantly are chosen to be reviewed less by developers.}
Fig.~\ref{fig:model} shows the relationship between the predicted level of dependency review and the closeness centrality measure.
The relationship between both variables for our first model is shown in Fig.~\ref{SUBFIG:ModelDifferent}.
The closeness centrality variable had the highest sum of squares, accounting for most of the model's results.

We find positive coefficients for the betweenness centrality and PageRank measures.
Hence, acting as an intermediary between contributors and engaging in highly-ranked collaborations within the network leads to less review from downstream. 
On the other hand, the eigenvector centrality measure has a negative coefficient.
Hence, collaborating with highly-ranked contributors in the network leads to a higher level of review from the downstream.
We hypothesize that higher-ranked contributors interact with many lower-ranked contributors, possibly leading to a negative relationship.
Still, all measures except closeness centrality only account for 25\% of the model's results, as indicated by the sum of squares.

The $R^2_m$ indicates that the fixed effects, contributor reputation measures, account for 6\% of the model variation.
Additionally, we find through the $R^2_c$ that the fixed and random effects account for 13\% of the model variation.
Therefore, we find that closeness centrality is a statistically significant variable for the chosen level of dependency review.
\textbf{Yet, developers consider more than contributor reputation when reviewing dependencies.}
Several reasons were mentioned in Section~\ref{sec:rq1}, and we discuss more about the implications and future work in Section~\ref{sec:discussion}.

The second model distributed 1477 responses in 795 groups.
Closeness centrality was also a positive and statistically significant fixed effect for downstream review, as shown in Fig.~\ref{SUBFIG:ModelSame}.
The sum of squares indicated that closeness centrality explained most of the results.
The betweenness centrality measure had a positive coefficient, yet the sum of squares was lower than the first model.
The eigenvector centrality also had a negative coefficient but was the second most significant variable in the model, as detailed by the sum of squares.
Interestingly, compared to the first model, PageRank centrality was a negative coefficient, indicating that participating in high-ranking collaborations in the network leads to higher review levels.
Still, PageRank centrality accounts for little of the model's results.
As all the answers respondents gave were the same for this model, due to reasons explained in Section~\ref{sec:rq1}, the goodness of fit was less than 0.01\% for both $R^2_m$ and $R^2_c$.
Closeness centrality as a model fixed-effect was, nonetheless, a statistically significant variable even when developers applied the same level of review without differing based on contributor identity.

The third model captured 2406 responses in 1170 clusters of responses. 
The model exhibits similar results to the second model but with some notable differences.
The statistical significance of the closeness centrality, shown in Fig.~\ref{SUBFIG:ModelAll}, is higher.
The $R^2_m$ and $R^2_c$ have higher values, corresponding to 2\% in both cases.
Lastly, the betweenness centrality measures account for more of the model variation as indicated by the sum of squares.

%% file: sections/6.discussion.tex
\section{Discussion}
\label{sec:discussion}
This section discusses our findings and recommendations.

\textbf{Ecosystems should provide a contributor reputation badge.}
We found that 51\% of developers consider reputation when adding or updating a dependency.
Similarly, prior work has also found developer-related factors relevant when developers review pull requests in their projects~\cite{tsay2014influence, zhang2022pull, gonzalez2021Anomalicious}. 
We recommend ecosystems like GitHub, Rust, and npm to incorporate contributor reputation signals to simplify adoption, standardize calculations, and reduce measurement tampering.

Ecosystems can calculate a contributor reputation score for each developer using the coefficient of our models with the centrality measures. 
Since closeness centrality is the only statistically significant measure, using the measure exclusively is sufficient. 
Proxying contributor reputation with centrality measures provides the following benefits. 
First, the proxy is grounded and established in prior work (Section~\ref{bac:dsn}).
Second, the measures can be harder to manipulate than other profile measures due to requiring collaboration activity.
For example, bots can increase the number of followers of an account.
Manipulating file co-edition collaboration requires repository write permissions or a merged pull request in repositories that force review. 
Developers may manipulate author-reviewer collaboration by creating irrelevant pull requests. 
We can account for such manipulation by only including accepted pull requests and rate-limiting the number of author-review collaborations in a window. 
However, future work needs to analyze the effect of the change in the measure. 

Integrating the ecosystem signal as a binary badge can communicate that a developer is reputed while reducing the adverse effects of assigning a score to each contributor.  
Social media platforms have employed similar badges for user accounts to reduce impersonation~\cite{xiao2023account}.
Future work can also create project or package ecosystem-level reputation badges.
Further research in human-computer interaction should explore the best and most ethical ways to present the badges to developers. 
We also leave as future work establishing the thresholds for determining who is a reputable contributor within the network.

\textbf{Developers should consider contributor reputation badges when reviewing dependencies.}
We found that when reviewing dependencies, most developers consider reputation.
We thus recommend that developers, even developers who do not commonly use contributor reputation, consider the badge during dependency review.
Through the badge, developers can determine at a glance if a change was done by a reputable contributor, providing benefits for developers in different roles.
First, downstream developers can quantify the degree of reputed contributors working on a dependency or a specific version.
Hence, dependents can prioritize which dependencies with their versions to review and adopt.
Second, upstream developers can determine if a pull request requires further review or if a developer should gain maintainer repository access.
Third, newcomers benefit from additional ecosystem information, as they may not yet be familiar with reputed contributors.

Contributor reputation signals may help developers safeguard against software supply chain attacks like the xz-backdoor~\cite{freund2024}.
As an attack vector, the backdoor required social aspects to become an xz-utils maintainer, achieved by pressuring the original maintainer~\cite{what2024, everything2024}.
A contributor reputation signal could determine whether the developer seeking maintainer access and the accounts that pressured the original maintainer were highly reputed.
A signal could have also indicated the change of maintainer reputation in the dependency for dependents.

\textbf{Researchers and companies should assist developers in reviewing dependencies.} 
We found that most respondents answered that they infrequently review dependencies.
Developers mentioned employing strategies to reduce their workload as reviewing all dependencies is daunting, including skimming updates and selecting dependencies deemed secure by them.
Survey respondents mentioned that they are not supported to review dependency changes.

However, someone must review dependencies to safeguard against vulnerable or malicious code.
The software supply chain industry should not delegate full responsibility to contributors.
Instead, we argue that research and companies should assist in reviewing dependencies. 
Researchers can help automate reviews to adopt and update dependencies. 
Our contributor reputation measure represents a step in this direction, though additional tooling and measures can help with this effort. 
For example, through recommendation systems for selecting and replacing dependencies~\cite{nguyen2020crossrec}, and commit anomaly detection~\cite{gonzalez2021Anomalicious}.
Meanwhile, companies can contribute back to the open-source community~\cite{wermke2023always} by sharing dependency audit results in platforms such as cargo-crev~\cite{cargocrev} and cargo-vet~\cite{cargovet}.
Companies can also support the security of critical open-source projects by supporting initiatives like OpenSSF Alpha-Omega~\cite{alphaomega}.
Helping review dependency updates is also required as security risks may emerge after a dependency is adopted and evolves as dependencies are continuously updated.

\textbf{Framework authors should incorporate contributor reputation badges.}
Practitioners have proposed framework solutions to safeguard against vulnerable or malicious commits (Section~\ref{bac:ssc}).
Authors of the frameworks can integrate our proposed ecosystem contributor reputation badge to strengthen security guarantees.
For example, SLSA~\cite{slsa} can indicate if two reputed contributors reviewed all code changes, reincorporating the highest security rating~\cite{slsaFuture}.
OpenSSF Scorecard~\cite{scorecard} can check if reputable contributors or maintainers reviewed or authored the latest changes. 
Dependabot~\cite{dependabot} can also monitor changes in the reputation of dependency maintainers.
Auditing platforms, such as cargo-crev~\cite{cargocrev} and cargo-vet~\cite{cargovet}, can indicate the dependency auditor's reputation.

\textbf{Researchers should analyze other signals for reviewing dependencies.}
Our results are promising for using contributor reputation as a signal to help prioritize dependency review efforts.
Yet, as our models and respondents indicated, contributor reputation does not fully account for how developers choose to review dependencies. 
A complete list of signals we found is available in the supplemental material~\cite{us2024Supplemental}.
Prior research has discovered similar signals used by developers to select projects to contribute to~\cite{qiu2019signals} and choose dependencies~\cite{miller2023We, holtervennhoff2023Wouldn, zahan2022weak}.
We leave analyzing other dependency review signals to future research.
As we find that developers vary their review when adding and updating a dependency, future work should account for the differences during study design.

%% file: sections/7.limitation.tex
\section{Limitations and Threats to Validity}
\label{limitation}

In this section, we discuss the limitations of the contributor reputation measure and the threats to validity.

\textbf{Limitations of the network centrality measure.} 
The network centrality measure as a proxy for contributor reputation has the following limitations.
First, the measure is no substitute for reviewing changes in dependencies.
Even if changes come from reputed contributors, anyone can make mistakes and accounts may be compromised.
Still, our measures help developers prioritize review efforts.
Second, calculating the measure at an ecosystem level is costly and may be limited by APIs.
The limitation can be reduced through our recommended ecosystem-provided contributor reputation badge.

\textbf{Construct validity.} 
We use network centrality measures as a proxy for contributor reputation.
Additionally, we operationalize developer collaboration through author-review and file co-edition activities.
The measures may not operationalize to our intended construct.
Though other operationalization can be explored in future research, we mitigate the threat as our operationalizations are grounded in prior work~\cite{bosu2014impact,meneely2011socio, kerzazi2016can}.

\textbf{Internal and conclusion validity.}
The network we constructed is limited to the data stored in Crates.io and GitHub that we could gather using the APIs.
Though we may not have captured all existing developer relationships, the threat is mitigated as prior work found developer social network metrics robust even with incomplete data~\cite{nia2010validity}.
As bot accounts may be present within the network, we mitigate the threat of including bots by discarding accounts using a heuristic~\footref{foot:bot}.
We acknowledge, as we did not alias GitHub accounts' emails, that we may be segmenting developer collaboration activity in our network.

Our survey findings may be subject to unknown biases based on who chose to participate in our survey. 
The threat is limited as our response rate of 14.3\% is in line with developer studies~\cite{smith2013improving}.
Our purposeful stratified sampling approach to select ten contributors for the survey may have introduced biases for respondents.
We deliberately chose the approach instead of a random sample to increase the chance of familiarity and the diversity of chosen contributors. 
To mitigate the threat, we only specified the list of GitHub users without the aggregate centrality score.
Two researchers designed the survey to reduce survey misunderstandings by asking objective questions.
Still, when analyzing the open-ended answers in our survey, we noticed that some respondents appeared to have misinterpreted our questions.
We reported all answers to reduce the introduction of additional biases when discarding responses.
We acknowledge the threat to the conclusions of our results.
Our qualitative analysis is subject to judgment.
To mitigate the threat, we analyzed the open-ended questions between two researchers in an iterative process to refine our codes until we achieved an inter-coder agreement of 0.82.

\textbf{External validity.}  
We acknowledge that our findings are limited to the Rust ecosystem, the top 1,644 packages we analyzed, and the time period selected.
To generalize our results, further research should be performed.

%% file: sections/8.conclusion.tex
\section{Conclusion}
\label{conclusion}

In this work, we analyze how contributor reputation can serve as a signal to help developers prioritize dependency review efforts.
Grounded in prior work, we proxy contributor reputation using network centrality measures.
We employ a mixed methods methodology within the Rust ecosystem to (i) construct a social network of collaboration activity, (ii) survey how developers choose to review dependencies, and (iii) model the effectiveness of centrality measures as a signal.
Among our findings, developers are not commonly (24\%) reviewing dependencies.
Often, developers depend on other strategies to not fully review dependencies, including considering contributor reputation (51\%).
Our regression models show that network centrality measures can serve as a signal for dependency review.
However, survey and model results indicate that contributor reputation measures alone do not fully account for how developers choose to review dependencies.
In our survey, developers mentioned other factors that should be studied in future work, including project popularity, dependency usage, and the project context.
Our main recommendation is for ecosystems such as GitHub, Rust and npm to provide a badge for contributor reputation.
Developers can leverage the badge to help in dependency reviews.

%% file: sections/9.acknowledgments.tex
\section*{Acknowledgments}

The work was supported and funded by the National Science Foundation Grant No. 2207008, North Carolina State University Provost Doctoral Fellowship, and Goodnight Doctoral Fellowship. 
Any opinions expressed in this material are those of the authors and do not necessarily reflect the views of any of the funding organizations. 
We thank the Realsearch and WSPR research groups from North Carolina State University for their support and feedback.